\documentstyle[11pt,aas2pp4]{article}
\input{psfig}
\input epsf 

\accepted{March 6, 1997}
\journalid{August 1, 1997}{484}
\begin{document}

\title{Skymapping with OSSE via the Mean Field
Annealing Pixon Technique}

\author{D. D. Dixon, T. O. T\"umer, and A. D. Zych}
\affil{University of California\\
        Institute for Geophysics and Planetary Physics\\
        Riverside, CA 92521}

\author{L. X. Cheng}
\affil{Astronomy Department\\
	Univ of Maryland\\
	College Park, MD 20742}

\author{W. N. Johnson and J. D. Kurfess}
\affil{E. O. Hulburt Center for Space Research\\
Naval Research Laboratory\\
Washington, DC 20375}

\author{R. K. Pi\~na and R. C. Puetter}
\affil{Center for Astrophysics and Space Science\\
University of California, San Diego\\
9500 Gilman Dr.\\
La Jolla, CA 92093-0111}

\author{W. R. Purcell}
\affil{Northwestern University\\
Department of Physics and Astronomy\\
2145 Sheridan Rd.\\
Evanston, IL 60208-3112}

\and

\author{Wm. A. Wheaton}
\affil{High Energy Astrophysics Group\\
        Jet Propulsion Laboratory, 169-327\\
        California Institute of Technology\\
        Pasadena, CA 91109}

\begin{abstract}
We present progress toward using scanned OSSE observations for mapping
and sky survey work.  To this end, we have developed
a technique for detecting pointlike sources of unknown number and
location, given that they appear in a background which is relatively
featureless or which can be modeled.
The technique, based on the newly developed {\em pixon}
concept and mean field annealing, is described, with sample
reconstructions of data from the OSSE Virgo Survey.  The results
demonstrate the capability of reconstructing source information without
any {\em a priori} information about the number and/or location of
pointlike sources in the field-of-view.
\end{abstract}


\section{Imaging with OSSE}
The Oriented Scintillation Spectrometer Experiment (OSSE) aboard the
Compton Gamma Ray Observatory (GRO) consists of four actively shielded
NaI(Tl)-CsI(Na) phoswich detectors~\cite{johnson93}.  Collimation is
provided by tungsten collimators, which define a $3.8^\circ\times 11.4^\circ$
(FWHM) field of view.  Each detector is mounted on an independent
single-axis pointing system, which allows for instrument pointings which
are offset from the spacecraft $Z$-axis.  Complex pointing plans for the
OSSE instrument may be effected via the programmable detector
orientation system.\\

The original functional intent of OSSE was to be a ``point-and-count''
spectrometer, i.e., the detector would be oriented toward a source of
interest and counts would be accumulated, with offset pointings providing
background estimates for the observation.  The
ability to program more complex pointing plans, however, raises the
possibility of using the non-uniform aperture response to our advantage.
By scanning the instrument in steps smaller than the aperture size, and
taking scans that overlap in at least two different directions, 
we can (in principle)
use the knowledge of the aperture function and the differential flux measured
between detectors to distinguish features at a substantially
better resolution than that implied by the aperture size.\\

A key goal of OSSE scanned observations is to perform sky-survey work,
and attempt to detect previously unknown point sources.  
A separate but related project is to map the
low energy galactic $\gamma$-ray emission.
However, the nature of OSSE scanned observations
presents some major difficulties to standard data inversion techniques.
Typically, the total number of observations is small ($O(100))$, each
with a fairly low signal-to-background ($\sim 0.1$\%).  From this,
we would like to construct a map of the flux in a rather larger
set of pixels ($O(1000)$), where the pixels are significantly smaller
than the aperture size.  This is obviously an impossible task for
``direct'' deconvolution or inversion, and while model fitting is
more feasible, it is also undesirable due to the bias introduced
by selection of a particular model.
We have thus developed a new approach which is highly effective at
detecting point sources in an unbiased manner, i.e., with no previous
knowledge of their number, location, or strength.  
This approach begins with the problem phrased as one of
deconvolution, and ultimately transforms it to one of model fitting.\\

\section{Direct Deconvolution}
OSSE is a {\em linear} instrument, in that the detected signal is a linear
function of the source intensities (this holds true for many types of
instruments).  Due to the linear nature of OSSE, it is perhaps most
``natural'' to deal with statistical fluctuations (Poisson noise) in the
data by use of {\em linear least squares} (LLSQ) techniques.  For the
purposes of this paper, it will be useful to distinguish between two
types of LLSQ problems, namely model fitting and deconvolution.
By LLSQ {\em model fitting\/} we mean 
the adjustment of the co-efficients in a given linear model to fit
a given set of count data (as described, e.g., by Wheaton {\em et al.} 
1995, and references therein).
An example would be fitting an OSSE data set to a model consisting of
several point sources at known locations, and a  linear background 
model with a finite (small) number of terms of known form.
{\em Deconvolution}, on the other hand, refers to situations in which
the model is not known, or perhaps has a known form but a large
(in principle infinite) number of terms.
An example of deconvolution would be fitting the 
same OSSE data for observations containing diffuse
sources, or with an unknown number of sources at unknown positions.\\

The essential distinction between the two 
is that the former problem involves solving
a finite set of linear equations, whereas the latter requires solution
of an {\em integral\/} equation, e.g.,
\begin{equation} \label{eq:integral}
\bar{D}(y) = \int E(y,x)I(x) dx,
\end{equation}
where $\bar{D}$ is the expected data, $I$ is the image intensity
distribution as a function of spatial coordinates $x$, and $E(y,x)$
is a known function, essentially the OSSE aperture function in our
context, relating the image space $X$ to the data space $Y$.
Solving for $I(x)$ given $D(y)$ and $E$ is a 
linear inverse problem, of a type common in astronomy.\\

Discrete LLSQ model fitting methods may be applied 
to the deconvolution problem if we approximate 
the linear integral equation~(\ref{eq:integral}) 
by a finite matrix equation:
\begin{equation}\label{eq:matrix}
   \bf \bar{D} = E I,
\end{equation}
where now 
$\bar{D}_i = \bar{D}(y_i)$, $E_{ij} = E(y_i,x_j)$, and $I_j = I(x_j)$.
Because of the noise (counting statistics, in the Poisson case),
$\bf\bar{D}$ is never observed directly.
The measured quantity is the observed counts $\bf D$, which will contain
Poisson noise.
In astronomical imaging applications, we may try a huge model
representing the cosmic source terms, consisting of a source at 
``every possible location'', that is, an array of pixels 
spaced closely enough to be a good approximation to $I(x)$ 
in the integral~(\ref{eq:integral}).
This approach we have previously called
{\em Direct Linear Algebraic
Deconvolution\/} (DLAD, Dixon {\em et al.} 1993).
Wheaton {\em et al.} 1995 show that the DLAD estimate
coincides with the Poisson Maximum Likelihood (ML) if the
pixel array is the same.
As DLAD involves solving linear rather than non-linear equations, it
is computationally efficient, a point which shall prove crucial in what
follows.\\

Two difficulties arise at once.
First, if the data space $Y$ is binned finely to represent the
integral faithfully, then for the resulting tiny bins 
the expected counts 
$\bar{D}_i = \bar{D}(y_i)$
in data bin $i$ may be small, so that older Poisson LLSQ 
(PLLSQ) model-fitting methods fail due to ``insufficient statistics''
(that is, for small numbers of counts, the Poisson distribution is
poorly approximated by a normal distribution).
Second, when the pixel array $x_j$ is chosen fine enough, 
then $\bf E$, the discretized matrix representation of $E(y,x)$,
becomes highly ill-conditioned or singular.
Effective solutions to the first difficulty have been discussed 
by Wheaton {\em et al.} 1995.
The key point is that the weights for the LLSQ equations should 
be chosen to be uncorrelated with the data ${\bf D}$ 
itself, effectively ruling out the commonly used approximation 
$\sigma_i \approx D_i^{\frac{1}{2}}$ 
(as opposed to $\sigma_i = \bar{D_i}^{\frac{1}{2}}$, which is of course exact).
Bearing this requirement in mind, it is possible to construct PLLSQ 
estimators which are unbiased and otherwise convenient for arbitrarily
low $\bar{D_i}$.
With a weighting matrix $\bf W$ so chosen, the ``normal equations''
(see references in Wheaton {\em et al.} 1995) are
\begin{equation} \label{eq:pllsq}
\bf  E^T W^2 E I = E^T W^2 D,
\end{equation}
which may be solved for $\bf I$, given $\bf D$, in the usual way.\\

The second difficulty has often been summarized by describing the
discretization of the linear inverse problem as ``ill posed''.
We distinguish two aspects of the situation:  (a) failure of 
the numerical mathematical problem, which is essentially an artifact,
due to the incapacity of simpler matrix inversion
algorithms when faced with ill-conditioned or singular problems
(especially {\em large\/} problems),
and (b) the violent anti-correlation which arises among near-by
pixels when the pixel size is $\le$ angular resolution of the 
instrument.
The algorithmic aspect of the problem is resolved by recourse to
singular value decomposition (SVD; see Press {\em et al.} 1986 for
a simple discussion) in the computation.
The anti-correlation among pixels is mathematically inescapable, 
due to the real
confusion among nearby points, which the instrument is unable to
resolve clearly.
The problem can be partly removed by making use of the
{\em physical\/} constraint of non-negativity of
the source fluxes, as previously discussed in connection with 
imaging from COMPTEL data
(Dixon {\em et al.} 1993, and Wheaton {\em et al.} 1993, and 
more generally by Dixon {\em et al.} 1996).
The non-negativity constraint has been enforced by a variant of the
Non-Negative Least Squares (NNLS) algorithm of Lawson and Hanson (1974),
and the resulting technique is termed {\em Constrained Linear Algebraic
Deconvolution} (CLAD)~\cite{dixon96}.\\

\begin{figure}[t]
\epsfxsize=3.0in \epsfbox{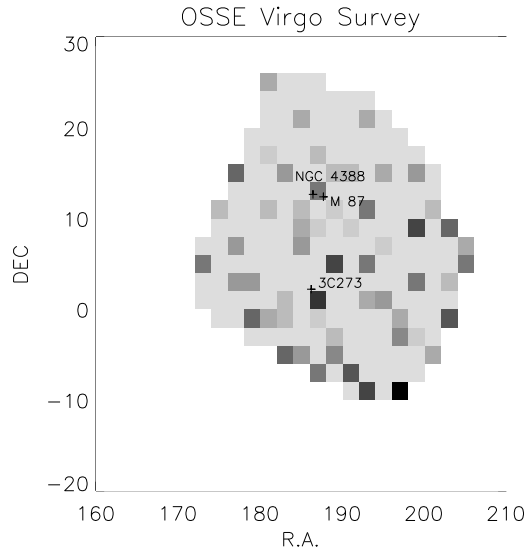} %[0 0 30 50]
%\caption{
\figcaption[fig1.ps]{
Image found generated by solving the DLAD equations, subject
to the constraint that the answer be non-negative.  Bright sources in
the FOV are labelled.  Note other bright, spurious sources.  The image
resolution is $2^\circ$.
\label{fig:nnls}
}
\end{figure}

Figure~\ref{fig:nnls} shows a map,
using data from the OSSE Virgo Survey,
made using the CLAD
method with a resolution of $2^\circ$ in the 50--150 keV energy band.
The irregularly shaped exterior region bounds the area 
of significant exposure during the scanned observation.
The exposure matrix $\bf E$ was calculated from
the OSSE aperture response, 
and gives the contribution to each data point from each sky pixel.
In this observation, the detectors were
stepped at increments of $\sim 1.8^\circ$ for individual scans, with
the spacecraft orientation changed $\sim 4^\circ$ between scans.
The two strong sources in the FOV, 3C 273 and NGC 4388, appear 
prominently, but there are several other spurious bright sources, 
albeit of lower statistical significance.  We denote these pixels
as ``spurious'' due to their lack of correspondence with any candidate
$\gamma$-ray source, as well as their marginal significance
(3C 279 is in the field, but no significant feature is
associated with it in the map).  An important point is that due
to the nature of the observations, the reconstructed source intensity
doesn't necessarily have a direct correspondence to its statistical
significance.  Note that M87 is within the
same pixel as NGC 4388, but it as yet unclear if its contribution to the
total flux is significant in this energy range.\\

While CLAD as illustrated in Figure~\ref{fig:nnls}
appears to be a useful tool, the spurious sources are, to say the
least, cosmetically unattractive.
That they are generally non-significant suggests that we could still
make a map which fits the data well even if they were somehow 
eliminated.
A more serious problem is that the large number of pixels retained
in the model seriously degrades the sensitivity of the maps
produced.
We are thus led to seek an objective means of eliminating from the 
model those pixels which are not statistically required by the data.
A development based on the {\em pixon\/} concept of Puetter and Pi\~{n}a
(1993), described in the following section, 
gives a powerful new image reconstruction algorithm which fulfills
this need.

\section{Pixon Based Inversion}

In a broad sense, the problems we encountered above occurred because
we tried to fit too many parameters, i.e., more than were statistically
justified by the data.  For example, consider a large region in an image
that is statistically consistent with being flat.  Ideally, we could
represent the information in this region with a single number, such as
the average flux.  However, since we don't necessarily know {\em a priori}
which regions are flat and which aren't, the standard approach is to
cover the FOV with small pixels, and find the flux in each.  In the
statistically flat region, where there is little coherent information
spatial information, but lots of noise, the inversion algorithm quite
happily places flux in many of the small pixels to achieve gains (albeit
statistically marginal) in the goodness-of-fit (GOF, e.g., $\chi^2$,
likelihood, etc.).  The problem, then, is less with the inversion algorithm
itself as opposed to the choice of image basis that we supplied to that
algorithm.\\

Many attempts to counteract this problem concentrate on placing some sort
of extra constraint (hard or soft) on the form of the solution, and in this
way, effectively reducing the number of ``free'' parameters.  A prime
example of this is Maximum Entropy (ME)~\cite{skilling89}.  
Such approaches are effective
to some degree, but often are limited in their realm of applicability,
as well as in that they yield biased solutions~\cite{donoho}.  The pixon
approach also attempts to reduce the number of free parameters, but does
so in a more direct and fundamental manner.
For the purposes of this paper, we shall limit
somewhat the discussion of the conceptual underpinnings of pixons;
for more details, the reader is referred to~\cite{puetter94}.  The basic
idea behind a pixon is that it is a sort of flexible pixel~\cite{puetter93},
able to change it's shape and size.  In the pixon approach, a large flat
area in the image would not represented by many pixels containing the same
intensity, but rather would be represented by a single pixon, with a
single number for the intensity and a few others describing the shape.
An image described by pixons would thus require a smaller set of parameters.\\

The criterion we use in this approach is 
essentially Occam's razor: the simplest
model which yields an image statistically consistent with the data is
the correct model.  The next question is how to go about {\em finding}
the simplest model.  The starting point is to define a 
{\em fundamental model}, consisting of a grid of pixels at the smallest
resolution we expect to be able to see, as well as the instrumental
response from those pixels.  For OSSE, this is represented by the
exposure matrix ${\bf E}$, whose columns form the fundamental basis.  All pixon
bases are simply derived as positive linear 
combinations of the fundamental basis
functions.  We represent this linear transformation by the matrix ${\bf K}$,
and thus the new model matrix is simply ${\bf E' = EK}$.  
Equation~\ref{eq:pllsq} then becomes
\begin{equation}\label{eq:pixon}
\bf K^T \alpha K I^{(p)}= K^T \beta,
\end{equation}
where ${\bf \alpha = E^T W^2 E}$ and ${\bf \beta = E^T W^2 D}$, and 
${\bf I^{(p)}}$ is
the {\em pseudo-image}.  Equation~\ref{eq:pixon} may be solved for
${\bf I^{(p)}}$ subject to the non-negativity constraint to yield the
constrained least-squares coefficients in the pixon basis.  To see the
actual sky map, we need the representation in the fundamental basis,
for which we simply compute ${\bf I = K I^{(p)}}$.\\

\begin{figure}[t]
\epsfxsize=3.0in \epsfbox{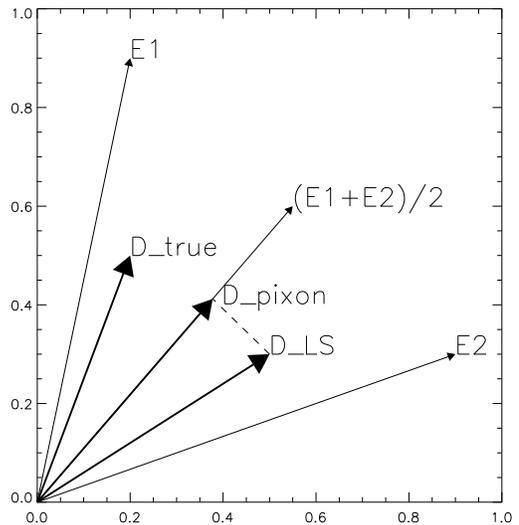} %[0 0 30 50]
%\caption{
\figcaption[fig2.ps]{
A schematic of the effect of the pixon prior.  $D_{true}$ is the
``noise-free data'', i.e. what we would measure in the limit of infinite
statistics.  $E_1$ and $E_2$ represent fundamental basis functions for
the instrument, while $(E_1 + E_2)/2$ is a basis function for a pixon basis.
$D_{LS}$ is the constrained least-squares estimate for the fundamental
basis, while $D_{pixon}$ is the constrained least-squares estimate for the
pixon basis.  Note that $D_{pixon}$ requires only one basis function as
opposed to two for $D_{LS}$, and further that $D_{pixon}$ is closer
to $D_{true}$.
\label{fig:pixexp}
}
\end{figure}

The reader at this point may wonder why we don't just substitute
${\bf I = K I^{(p)}}$ into eqn.~\ref{eq:pixon} and solve for 
${\bf I}$ directly.
For that matter, a quick bit of algebra would seem to indicate that
all of the ${\bf K}$'s cancel out when solving for ${\bf I}$.  
However, it is to
be noted that direct algebraic solution for ${\bf I^{(p)}}$ gives the
{\em unconstrained} least-squares solution.  Enforcement of the
non-negativity constraint implies that some elements of ${\bf I^{(p)}}$
are constrained to be zero.  Only those columns of ${\bf E'}$ which
will yield positive coefficients are used as the basis for the fit,
which proves to be the key in making the pixon approach effective.
Let us illustrate with a simple problem.  We take ${\bf I}$ 
to be two-dimensional
and ${\bf D}$ to be three-dimensional.  ${\bf E}$ is thus a $3\times2$ matrix,
which we represent as ${\bf [E_1, E_2]}$, where ${\bf E_1}$ and 
${\bf E_2}$ are the
3-vectors forming the fundamental basis.  Figure~\ref{fig:pixexp}
shows the plane spanned by the fundamental basis, where ${\bf D_{true}}$
is the true value of ${\bf D}$, which we would measure in the limit of
infinite statistics.  Measurement noise, however, will cause ${\bf D}$ to
be found away from ${\bf D_{true}}$, and generally
contain some component perpendicular to the plane.  So we perform
a least-squares fit to find ${\bf D_{LS}}$, which is simply the orthogonal
projection of ${\bf D}$ onto the plane.  As ${\bf D_{LS}}$ lies 
within the convex region
bounded by ${\bf E_1}$ and ${\bf E_2}$, the components of 
${\bf I}$ will be positive.
Now we form the pixon basis ${\bf E' = [E_1, E_1 + E_2]}$, and see that
${\bf D_{LS}}$ lies outside the convex region bounded by this basis.
The non-negative least-squares estimate ${\bf D_{pixon}}$ is found by 
projection
of ${\bf D_{LS}}$ onto the nearest positive subspace~\cite{werner90}, which in
this case is the vector ${\bf E_1 + E_2}$.  The pseudo-image is thus
${\bf I^{(p)} = [0, I^{(p)}_2]}$, giving the image 
${\bf I = [I^{(p)}_2, I^{(p)}_2]}$,
which is obviously not the same is the image corresponding to 
${\bf D_{LS}}$.\\

The above exercise, while rather simple, does show precisely why the
pixon approach is superior to straightforward constrained ML fitting.
The pixon basis, as it is formed of
positive linear combinations of fundamental basis functions, will
{\em always} bound a smaller convex region than the fundamental basis.
As shown in Figure~\ref{fig:pixexp}, if this reduced region contains
${\bf D_{true}}$, then the estimate ${\bf D_{pixon}}$ is likely to be 
closer to
${\bf D_{true}}$ than the ML estimate.  Occam's razor
has served us well: the fundamental model required two basis functions
to perform the fit, but due to the constraint, the pixon model required
only one, and yielded a superior estimate.  In the context of image
reconstruction, Occam's razor implies that the optimal pixon basis
will allow only that level of detail that is statistically justified
by the data.  Areas in the image which are statistically smooth should
be constrained to be smooth in the fit, thus reducing the number of
free parameters.  As a result, spurious sources are largely done
away with, and correspondingly, the flux estimates and sensitivity for
actual sources are improved.\\

\section{Finding the optimal pixon basis}

While the advantages of using the optimal pixon basis for image reconstruction
are obvious, what is not obvious is how one goes about finding it.
The set of possible basis functions consists of {\em all} positive linear
combinations of the fundamental basis functions.  Further, there exists
no {\em a priori} systematic way of ordering the possible ways of
choosing a basis, i.e., the only way compare the merits of two different
bases is to actually perform the fit for each one.  We must find some way of
systematizing the search procedure.\\

The first step is to choose some set of potential pixon shapes.  This 
eliminates the possibility of being absolutely optimal, but greatly shrinks 
the set of possible basis functions.  Following Puetter and 
Pi\~na~\cite{puetter93}, we have implemented a {\em fuzzy pixon basis}.
Here, the pixon boundaries are not sharply defined, and the pixons are
allowed to overlap.  For simplicity, we shall use pixons which are circularly
symmetric, and thus describable by a single parameter.
For the pixon shape we have chosen, there will be associated with each pixel 
in the image a length
$\delta_j$, called the pixon scale,
which defines the size of the pixon.
The pixon shape function we have chosen is an inverted
paraboloid~\cite{puetter94}, and thus the pixon transform matrix ${\bf K}$ 
is defined as
\begin{eqnarray}
F_{jk} & = & \left\{
\begin{array}{ll}
1 - \frac{d_{jk}^2}{\delta_j^2}, & d_{jk} \leq \delta_j,\\
0, & d_{jk} > \delta_j,
\end{array}\right.\nonumber\\
V & = & \sum_j F_{jk},\nonumber\\
K_{jk} & = & \frac{F_{jk}}{V},
\end{eqnarray}
where $d_{jk}$ is the distance between image pixel $j$ and pseudo-image
pixel $k$, and we have a one-to-one correspondence between the image
and pseudo-image pixels.  Further, we restrict $\delta_j$ to some
finite set of values.  The implementation used for this paper utilizes 
two values, a small one $\delta^{(1)}$ (usually one pixel) for pointlike
sources,  and a large one $\delta^{(2)}$ (roughly the size of the image space)
for modeling diffuse background emission.\\

Given the above restrictions on $K$, we now have a well-defined set of 
potential basis functions.  However, this set is still forbiddingly
large.  With two pixon scales and an image space of dimension $J$, the
number of possible bases is $2^J$, thus a brute force search is infeasible
for all but very small problems.  The problem of finding an optimal pixon
basis is thus one of combinatorial (rather than functional) minimization.
Practical techniques for attacking such problems are generally heuristic
in nature, and promise only a near-optimal solution.  For OSSE sky
survey work, we anticipate looking for point sources within an otherwise
roughly uniform background, implying that we really only need two pixon
sizes, one small (for point sources) and one large (for the background).
The choice of two disparate types of basis functions leads naturally to
an optimization heuristic
called {\em mean field annealing} (MFA)~\cite{reeves93}.  Annealing
algorithms derive from the statistical mechanical analog of annealing
a material from an initially disordered state, where one begins at a
high temperature and slowly cools the material.  Consider the case of
a ferromagnetic material.  In the presence of a uniform external field,
each atom in the material can have one of two possible spins (up or down), 
and the configuration space of the magnet is simply all possible 
combinations of ups and downs.
The configuration corresponding to a defect-free
magnet occurs at the global minimum of the Hamiltonian, where all of the 
spins point in the same direction.  However,
the Hamiltonian also has many local minima, corresponding to the formation
of domains, where groups of spins in different regions of the magnet point
in different directions.  To anneal a magnet, one begins at a temperature
high enough that $kT$ greatly outweighs the interaction energy.  At high
temperature, the spins will be oriented randomly, and on average half will
be up and half will be down.  If the system is cooled
too quickly, it ``quenches'', becoming trapped in a local minimum.  
Slow cooling, however,
allows the system to ``jump out'' of local minima via thermal fluctuations.
By lowering the temperature in steps, and allowing equilibrium to be
reached at each step, we can arrive at a nearly defect-free magnet
in the low temperature limit.\\

The problem of finding the pixon basis is similar.  Here, the configuration
space is all possible assignments of pixon sizes to the pixels of the
fundamental basis.  With two possible pixon sizes, the problem is closely
analogous to that of the magnet, where we tried to assign one of two
possible values of spin.  The Hamiltonian we choose consists of two parts:
the $\chi^2$, and a function $\sum_j s_j$, where $s_j$
is a linear function of $\delta_j$,
and is taken to be 1 at the smallest value of $\delta_j$, and
0 for the largest value.  This second function acts as a penalty against
complexity, utilizing the fact that smaller pixons (more detail) generally 
imply more parameters in the model.  The particular functional relationship of
$s_j$ to $\delta_j$ isn't so important, for as we'll see below,
it is only {\em changes} in $s_j$ that will be relevant.
We then write the following Hamiltonian for the problem:
\begin{equation}\label{eq:hamiltonian}
H(\delta) = \gamma\chi^2 + \sum_j s_j,
\end{equation}
with $\chi^2$ evaluated for the result of a fit using a particular set of
pixon scales $\delta$.  This Hamiltonian represents a balance between 
finding the simplest
model (all big pixons) and satisfying the $\chi^2$, with the constant
$\gamma$ weighting the relative importance of the two terms.
We then define a parameter $T$ analogous to temperature, and write
the Boltzmann distribution
\begin{equation}\label{eq:boltz}
p(\delta,T) = \frac{e^{-H(\delta)/T}}{Z},
\end{equation}
where $Z$ is the partition function, $Z = \sum_\delta e^{-H(\delta)/T}$,
with the sum performed over all possible configurations $\delta$.\\
  
Our ultimate goal is to find the average values $<\delta>$ at a given
temperature for the specified Hamiltonian and Boltzmann distribution.
Clearly, this calculation cannot be carried out exactly for an interesting
number of pixels, as it requires calculating eqn.~\ref{eq:boltz} over
$2^J$ configurations.  We note
that for $T$ constant, the equilibrium energy is equal to the average energy.
Define $H_{ij}$ as the average energies for states with $\delta_j$ held fixed 
at one of the two allowed values $\delta^{(i)}$, i.e., 
$H_{ij} = <H(\delta)>|_{\delta_j = \delta^{(i)}}$.  We then approximate
the average value of
$\delta_j$ at equilibrium for a given temperature 
eqn.~\ref{eq:boltz} as
\begin{eqnarray}\label{eq:dbar}
<\delta_j> 
& = & \frac{\delta^{(1)} \exp(-H_{1j}/T) + \delta^{(2)} \exp(-H_{2j}/T)}
			{\exp(-H_{1j}/T) + \exp(-H_{2j}/T)}\nonumber\\
& = & \frac{\delta^{(1)} + \delta^{(2)} \exp{[(H_{1j}-H_{2j})/T]}}
	{1 + \exp{[(H_{1j}-H_{2j})/T]}}.
\end{eqnarray}
where the mean or effective field
at $j$ is $\Phi_j = H_{1j}-H_{2j}$.
This seems easy enough.  However, calculation of $H_{ij}$ still requires
summation over {\em all} possible configurations.  
We thus make the additional approximation~\cite{vandenbout90}, 
that $<H(\delta)> \simeq H(<\delta>)$, i.e., the average value of the
energy is equal to the Hamiltonian evaluated at the average values
of $\delta$ (for the reader who finds this procedure a bit ad hoc and
suspicious, we shall investigate it more deeply in the following section).  
At each step in the algorithm, then,
$<\delta_j>$ is calculated from $H(<\delta>)|_{\delta_j = \delta^{(i)}}$,
with the values in $<\delta>$ being taken from the previous step,
leading to the MFA algorithm~\cite{vandenbout90}:
\begin{enumerate}
\item{$T = T_0 \equiv$ initial temperature}
\item{do until saturation
\begin{enumerate}
\item{do until equilibrium
\begin{enumerate}
\item{choose a random pixel $j$}
\item{set $\delta_j = \delta^{(1)}$}
\item{find the CLAD solution for this basis}
\item{Compute $H_{1j}$ for that solution}
\item{set $\delta_j = \delta^{(2)}$}
\item{find the CLAD solution for this basis}
\item{Compute $H_{2j}$ for that solution}
\item{compute $<\delta_j>$ from eqn.~\ref{eq:dbar}}
\end{enumerate}
}
\item{$T = cT$}
\end{enumerate}
}
\end{enumerate}
``Equilibrium'' is defined as the point where the $\delta_j$'s are no longer
changing significantly at a particular temperature (in practice, this
seems to require only one or two iterations).
``Saturation'' is the point where the $\delta_j$'s all take on
one of the possible predefined values, in this case, $\delta^{(1)}$ or
$\delta^{(2)}$.
The constant $c$ represents the cooling rate, and must be chosen
carefully.  If it is too small, the system quenches, and the low
temperature limit produces a solution that is in a non-optimal local
minimum.  If $c$ is too large, the algorithm takes too long to complete.
Nominally, the initial temperature may be chosen such that the system 
is initially
maximally disordered, i.e., $\delta_j = (\delta^{(1)} + \delta^{(1)})/2$.
In practice, one can choose a somewhat smaller temperature, where some degree
of structure (though not too much) is present in the $\delta_j$'s.
The constant $\gamma$ is also of importance, as overweighting of the
$\chi^2$ gives spurious sources, and underweighting suppresses weak
sources.  Note also from eqn.~\ref{eq:dbar} that we only use the
{\em difference} between $H_{1j}$ and $H_{2j}$.  Since we modify only
the $\delta_j$ for a particular $j$ while holding the others constant,
$H_{1j} - H_{2j} = \Delta \chi^2 - \Delta s_j$.  Given the definition
of $s_j$, $\Delta s_j$ is simply 1.\\

We note that this algorithm requires a good deal of solving
of the least squares problem for different bases, which may seem
to impose a large computational burden.  However, the variant of NNLS
we have developed mitigates this to a great extent.  Partially, this
is because we are solving the least squares problem within a linear
algebraic context, which is much more computationally efficient than
non-linear minimization.  Another contributing factor is our variant
of NNLS, which takes advantage of the fact that only some subset of the
parameters is unconstrained, and uses a $QR$ factorization updating 
scheme rather than re-solving the entire problem when only a single 
$\delta_j$ is changed.  This algorithm will be described in a
subsequent paper; a brief description is given here.
The $QR$-decomposition algorithm factors a matrix into an orthonormal
part ($Q$) and an upper triangular part ($R$)~\cite{gill}.
A useful property of $QR$ factorization is that rank one updates
(e.g., addition of a column to the original matrix) are relatively
easy to perform~\cite{gill}.  The original Lawson and Hanson implementation
of NNLS takes advantage of this~\cite{lawson74}.  Within the pixon
framework, addition/removal of a particular pixon to/from the active
set is also a rank one update.  Further, this action generally doesn't
modify the constraint set a great deal.  By using the results from
a previous NNLS step, and applying the $QR$-updating, we generally
save a good deal of time over redoing the entire NNLS procedure.
The result is that for $J$ pixels, $N$ possible pixon
sizes, $L$ temperature steps, and an average of $M$ unconstrained parameters,
the algorithm performs like $O(J L N M^2)$.\\

\section{Further discussion of the algorithm}
In this section, we present a more detailed look at the inner workings of
the mean field annealing algorithm described above.  The ``proof is in the
pudding'' reader may wish to skip this section on first reading, and move
directly to the section on the application of the algorithm to OSSE data.\\

The above derivation of the mean field equations (following that 
in~\cite{vandenbout90}), while based on seemingly reasonable assumptions,
does not make much rigorous contact with statistical mechanics.  To make
this connection clear, we will present the derivation given
in~\cite{reeves93}, which is a standard mean field approximation procedure.
We begin with a set of $J$ ``spins'', $s$, which are allowed to take on the
values 0 and 1.  The interactions between these spins is described by
a Hamiltonian or energy function, $H(s)$.  Correspondingly, for a system
which is Boltzmann distributed at temperature $T$, 
the partition function $Z$ is given by
\begin{equation}
Z = \sum_{[s]} e^{-H(s)/T},
\end{equation}
where the sum is taken over all possible configurations of $s$.  Next,
we replace $s$ with continuous variables $v$:
\begin{equation}
Z = \sum_{[s]} \int d[v] e^{-H(v)/T} \prod_j \delta(s_j - v_j),
\end{equation}
which, via the presence of the $\delta$-function, is the equal to the
previous expression.  The function $H(v)$ now represents an effective
energy in terms of the new variables.
We then Fourier expand the $\delta$-functions
in conjugate variables $u$, obtaining
\begin{equation}
Z = \sum_{[s]} \int d[v] \int d[u] e^{-H(v)/T}
\prod_j e^{u_i(s_j - v_j)},
\end{equation}
and carry out the sum over $s$:
\begin{equation}
Z \propto \int d[v] \int d[u] e^{-H(v)/T - \sum_j u_j v_j
+ \sum_j \log(1 + e^{u_j})}.
\end{equation}
We have thus rewritten the partition function entirely in terms of the
new variables, and as yet have made no approximation.  Next,
the saddlepoint approximation is made, 
which approximates the value of the integrand
by it's maximum value.  The point where the integrand takes on its
maximum value is given by
\begin{eqnarray}
u_j & = & -\frac{1}{T}\frac{\partial H(v)}{\partial v_j},\nonumber\\
v_j & = & \frac{1}{1 + e^{u_j}}.
\end{eqnarray}
Combining these gives the implicit equation for $v_j$:
\begin{equation}\label{eq:meanfield}
v_j = \left(1 + e^{-\frac{1}{T}\frac{\partial H(v)}{\partial v_j}}
\right)^{-1}.
\end{equation}\\

The question immediately arises as to the interpretation of the new
variables $v$, for at first glance, eqn.~\ref{eq:meanfield} bears little
resemblance to something like eqn.~\ref{eq:dbar}.  For many optimization
problems, such as graph partitioning~\cite{vandenbout90}, the form of
$H$ is such that $\partial H/\partial v_j = \Phi_j$, the mean field at
spin $j$.  In this case, $v_j$ in eqn.~\ref{eq:meanfield} takes the form
of a thermal average of $s_j$, evaluated in the mean field approximation.
The Hamiltonian in our application, however, is not so simple.  In this
case, the exponent in eqn.~\ref{eq:meanfield} is given by
\begin{equation}
\gamma\frac{\partial \chi^2}{\partial v_j} + 1,
\end{equation}
where the functional dependence of $\chi^2$ on $v$ is given via the
relation $\delta_j = \delta^{(1)}(1 - v_j) + \delta^{(2)}v_j$.  Given
that we evaluate $\chi^2$ {\em after} solving the constrained least
squares problem in the basis associated with $\delta$, it is not at
all clear that one can obtain a functional form for its partial derivatives.
Even if we were to make the reasonable assumption that small perturbations
in $\delta$ do not change the constraint set, the expression for the
partial derivative is quite complicated, and difficult to evaluate
numerically.  An obvious remedy is to make a numerical approximation
of the derivative,
\begin{equation}
\frac{\partial \chi^2}{\partial v_j} \simeq 
\frac{\Delta \chi^2}{\Delta v_j},
\end{equation}
with $\Delta v_j = v_j^{(2)} - v_j^{(1)}$, and $\Delta \chi^2$ the difference
in $\chi^2$ values evaluated at $v_j^{(2)}$ and $v_j^{(1)}$.  Substituting
the approximate derivative into eqn.~\ref{eq:meanfield}, setting
$\Delta v_j = 1$, and doing a little algebra, we obtain
\begin{equation}
v_j = \frac{e^{-H_{2j}}}{e^{-H_{1j}} + e^{-H_{2j}}},
\end{equation}
In this expression, the interpretation of $v_j$ as the thermal average
of $s_j$ within the above approximations, is obvious, and as such, leads
directly to eqn.~\ref{eq:dbar}.\\

Another question which arises is whether or not the mean field approximation
is even valid for the problem at hand.  Methods exist in statistical
mechanics for deciding this~\cite{pfeuty}, but are difficult to apply in
a general manner, given the nature of $H$ in our problem.  Generally
speaking, one expects the approximation to be reasonable in the case
where the number of ``spins'' is large, so that individual interactions
are small compared to the mean field, and/or when the dimensionality of
the system is large, such that each ``spin'' has many neighbors.  
For the image reconstruction problem, we would
expect the first condition to be satisfied when the FOV is a fair amount
larger than the extent of the instrument response, such that there exist
many groups of pixels which are effectively decoupled.  The second
condition is more subtle.  The dimensionality of our system is nominally
2, but, when we are attempting to superresolve the image, i.e., the pixel
size is smaller than the extent of the instrument response, the ``forces''
in the problem are effectively long range, as spatially separated pixels
will be correlated via the response.  For long range forces, the
effective dimension of the system is larger than the geometric
dimension~\cite{pfeuty}.  The precise calculation of this number for the
image reconstruction problem is difficult, and problem dependent.  Regardless,
it should be clear that for the case of small pixels, each ``spin'' will
feel the influence of many nearby ``spins''.\\

Now, we have obviously made just about the coarsest approximation possible
for $\partial H/\partial v_j$, but there are some decided benefits to
the particular choice we have made.  First, as we saw above, it allows
us to recover the interpretation of $v$ as an approximate thermal average.
Closely related is the property that for the approximation we have chosen,
as $T \rightarrow 0$, $v_j$ (and thus $<\delta_j>$) takes on one of its two
extremal values, which was really our goal in the first place.  
Due to the complicated nature of $H(v)$, this won't
necessarily occur for the exact solution of eqn.~\ref{eq:meanfield} in
the low $T$ limit.  This property is of some importance in sky survey
work, where we don't really want sources showing up as different sized
pixons depending on the statistics of the dataset.\\

The final benefit has to do with the method of solving for $v_j$.  As
eqn.~\ref{eq:meanfield} is implicit in $v_j$, direct solution is impossible.
A simple way of solving for $v_j$ is that of the algorithm described
above, which in terms of eqn.~\ref{eq:meanfield} is just the iteration
\begin{equation}\label{eq:mfd}
v_j^{(n+1)} = 
\left(1 + e^{-\frac{1}{T}\frac{\partial H(v^{(n)})}{\partial v_j^{(n)}}}
\right)^{-1}
 = f(v^{(n)}).
\end{equation}
The solution of eqn.~\ref{eq:meanfield} is represented by a fixed point
(of which there is likely more than one) of the mapping defined in 
eqn.~\ref{eq:mfd}.  Now, as $f(v^{(n)})$ of eqn.~\ref{eq:mfd} necessarily
lies between 0 and 1, we are guaranteed the existence of a fixed point.
We are {\em not}, however, guaranteed of convergence to that fixed point.
The stability of the fixed point is basically determined by the eigenvalues
of the Jacobian of the map in the neighborhood of the fixed point
(actually, the analysis is somewhat more complicated in the serial updating
scheme which we are using, but the flavor is the same~\cite{reeves93}).
If the eigenvalues $\lambda_i$ are such that $-1 < \lambda_i <1$, the
fixed point will be stable, and we will have convergence.  If not, other
undesirable behavior will occur.  Though a full eigenvalue analysis is
clearly out of the question, numerical experiments have shown that using
``better'' approximations for $\partial H/\partial v_j$ will often result
in behavior which is oscillatory, or even chaotic.  In our coarse
approximation, however, $\Delta v_j = 1$, and $0 \leq f(v) \leq 1$,
so $\partial v_j^{(n+1)}/\partial v_j^{(n)} \leq 1$.  Though we don't
prove that this implies stability of the fixed point, we do note that the
nature of the problem is such that the Jacobian will likely be diagonally
dominated (i.e., pixons which are well separated have little effect on
each other w.r.t. $v$ or $\delta$), and so we would expect our coarse
approximation to have some stabilizing effect; and in fact, in all of
our applications of the algorithm, convergence has always been achieved
using this approximation.\\

Though we have argued for the proposed algorithm's effectiveness in the
case of {\em two} pixon sizes, it should be fairly clear that it won't
be very effective for three or more.  In the above, we see that $H(v)$
is essentially replaced by a linear approximation, which is fine for
capturing the average behavior between two extremal pixon sizes.  However,
with an intermediate pixon size, one expects that, in some cases, $H$ will
have a significant minimum in between the two extremal values, and the
linear approximation fails to capture this.  Thus the extension of the
MFA algorithm to $>2$ spin states as suggested in~\cite{vandenbout90}
is not appropriate for the problem at hand.  We suggest that $>2$
pixon sizes might be accomodated within the MFA framework via a higher
order approximation for $H$, but leave this as an avenue for future
research.\\

\section{Application to OSSE data}
\begin{figure}[t]
\epsfxsize=3.0in \epsfbox{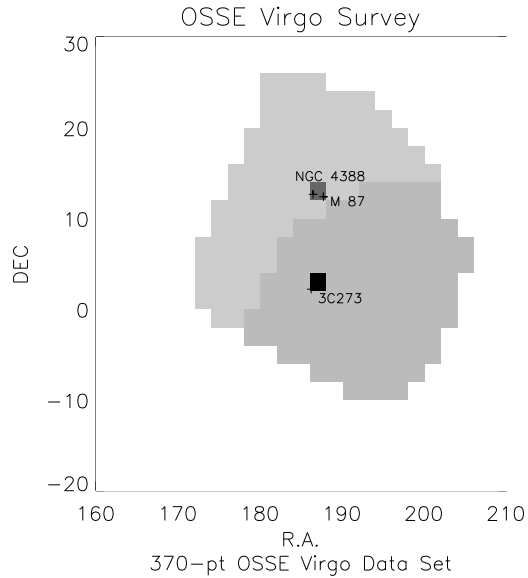} %[0 0 30 50]
%\caption{
\figcaption[fig3.ps]{
Image generated using the MFA pixon algorithm to find a
near-optimal pixon basis.  Note that the strong sources stand out, but
that the spurious peaks common in Figure 1 have been
suppressed.  The image resolution is $2^\circ$.
\label{fig:pixon}
}
\end{figure}

\begin{figure}[t]
\epsfxsize=3.0in \epsfbox{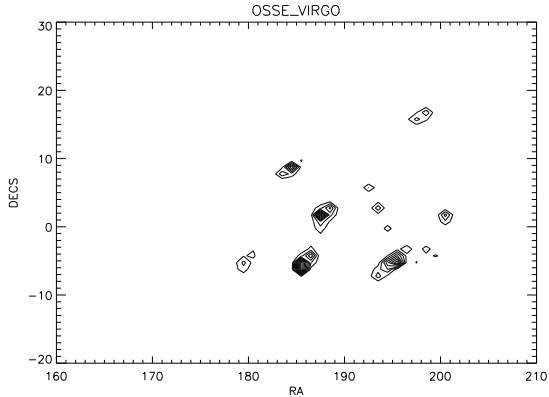} %[0 0 30 50]
%\caption{
\figcaption[fig4.ps]{
The same data as in Figure 3 reconstructed with Maximum Entropy.
Apparent sources other than those which are known to occupy the region
are of marginal significance, and we attribute them to the scan dependent
background effect, which appears as a broad gradient in Figure 3.
\label{fig:maxent}
}
\end{figure}

To test the effectiveness of our MFA pixon algorithm, we applied it to
the same dataset used to compute Figure~\ref{fig:nnls}.  For this purpose,
we set the initial temperature $T_0 = 3.$, $\gamma = 0.01$, and
$c = 0.3$.  The computation required approximately 15 minutes of CPU time
on a VAXstation model 4000-90 to compute the image for 209 pixels.
The result is shown in Figure~\ref{fig:pixon}, and is obviously
far superior to the constrained least squares result.  Most notably,
{\em none} of the spurious sources seen in Figure~\ref{fig:nnls} are
found by the MFA algorithm, and the two expected sources stand out clearly.
An apparent broad gradient is also noticeable; this is a known background
effect for OSSE scanned observations, which we have intentionally not
removed to show the utility of the pixon approach.
For comparison with a more established technique, Figure~\ref{fig:maxent}
shows a Maximum Entropy reconstruction of the same dataset.
The suppression of 
spurious sources by the MFA pixon algorithm is not only aesthetically pleasing,
but also implies that more confidence can be placed in the flux estimates
of the real sources.
Another measure of the success of the MFA algorithm is the number of
basis functions required by the fit.  The CLAD fit, using
the NNLS algorithm, required 56 basis vectors.  The pixon image required
only 4. 
It should be emphasized that the image in Figure~\ref{fig:pixon} was
generated without any {\em a priori} bias given to those pixels
containing real sources.  All pixels are treated equally by our
algorithm, with no assumptions being made about source locations.
Further results for the Virgo survey are presented in a separate
paper~\cite{kurfess95}.\\
  
\section{Future directions}
The MFA pixon algorithm proves to be remarkably successful, despite certain
limitations which will be addressed by future research.  
Most notably, there currently
exists no rigorous way to estimate the parameters $T_0$, $c$, and
$\gamma$.  For application to OSSE data, we have made a heuristic
estimate of $\gamma = 1/\Delta\chi^2(3\sigma)$, where 
$\Delta\chi^2(3\sigma)$ is the change in the $\chi^2$ required to
produce a $3\sigma$ change w.r.t. the $\chi^2$ distribution where the
number of degrees of freedom is calculated from the number of data points.
The idea is that any addition of further detail to the model which
produces less than a $3\sigma$ improvement to the $\chi^2$ will be
outweighed by the pixon penalty of $\Delta s_j = 1$.
From there, we may further refine our estimate of $\gamma$ by examination
of the final $\chi^2$ value.  If the $\chi^2$ falls outside the desired
confidence interval, $\gamma$ should be lowered to admit more parameters in
the final model.
The other two parameters, $T_0$ and $c$ were essentially guessed.
One would like to minimize $T_0$, so as not to spend time computing
iterations which produce no structure in the image.  By examining results
from the first iteration, one can tell if $T_0$ was chosen too large or
too small.  If too large, the $\delta_j$'s will show no structure, all
having essentially the same value.  If too small, some of the $\delta_j$'s
will saturate to one of the predefined values.  Somewhere in between these
two extremes lies the correct $T_0$, where some (but not too much) structure
is apparent, but at this point, judgements of how much structure is
appropriate are basically subjective.  Similarly, a small value of $c$
will hasten the convergence of the algorithm, but a value too small
will lead to a solution in a non-optimal local minimum.  For both
$T_0$ and $c$, it is best to err on the side of caution and
{\em overestimate} their values.  This leads to a somewhat larger
computational burden, but also increases confidence that the near-global
optimum will be found.\\

Despite this seemingly large amount of arbitrariness, it turns out that
the results from the MFA pixon algorithm are essentially constant over
a large range of parameters.  Experience has shown that even quite
bad estimates of $T_0$, $c$, or $\gamma$ still lead to excellent results.
We have changed the weighting factor $\gamma$ over as much as two orders
of magnitude, and at worst have seen only a few low significance spurious
sources, which still represents a vast improvement over previous methods.
Local minima obtained when $T_0$ or $c$ are underestimated generally show
sources shifted by one or two pixels, but statistically important detail
is never omitted completely (provided $\gamma$ isn't horribly 
underestimated).\\

\section{Conclusions}

The ability to create skymaps using data from OSSE scanned observations
has been demonstrated.  For this purpose, we have generated a new
image reconstruction algorithm which utilizes the concept of the
{\em pixon}.  The driving force behind the pixon concept is simply 
Occam's razor,
where we attempt to find the simplest possible basis for the reconstruction
that yields an acceptable value for the $\chi^2$.  As the problem of finding
such a basis is combinatoric in nature, we have employed an approximate
technique, mean field annealing, which gives a near optimal solution
with reasonable computational requirements.  The MFA pixon algorithm
has been successfully applied to scanned data from the OSSE Virgo Survey,
producing skymaps at resolution substantially smaller than the size
of the OSSE aperture, while suppressing spurious sources common in
pure maximum likelihood reconstructions.  Further, our algorithm has
been shown to find the nearly minimum basis for the reconstruction, 
implying smaller errors on source flux estimates.  The algorithm
is most effective for the case where the source distribution consists
of some number of highly localized sources in an otherwise slowly
varying ``background''.  We see that the MFA
pixon algorithm occupies a place somewhere between standard deconvolution
algorithms and model fitting algorithms, as it essentially transforms
the former problem into the latter.\\

One final note: the algorithm we have described has a somewhat different
flavor than commonly used regularized inversion schemes.  Techniques
such as Maximum Entropy estimate the {\em image} by finding the minimum
of the sum of two (or more) functions of the image parameters.  One
of these functions specifies the goodness-of-fit, while the other tends
to have the property of suppressing oscillatory behavior in the solution.
The MFA pixon algorithm, however, clearly is {\em not} estimating the
image, but rather $\delta$.  In this case, the $\chi^2$ plays the role
of a constraint, with $\gamma$ acting as a Lagrange multiplier.  From
these $\delta$, the image is obtained via a simple constrained least
squares estimate.  So, despite the fact that it appears that we've
reduced the number of parameters in the problem, we've actually just
changed the parameter set.  The effectiveness of the technique is at least
due in part to the fact that between the $\chi^2$ and the non-negativity
requirement on the image, that the total number of constraints is
large, and the problem thus highly overdetermined.\\

\begin{acknowledgements}
DDD wishes to thank J. Wudka for helpful discussions concerning mean field
theory.
Part of the work described in this paper was carried out by the Jet Propulsion
Laboratory, California Institute of Technology,
under contract with the National
Aeronautics and Space Administration.
This work was funded by NASA CGRO Grants 5-2044 and 5-2825.
\end{acknowledgements}

\end{document}